\documentclass[prb,aps,preprint]{revtex4}
\usepackage{amssymb}
\usepackage{amsbsy}
\usepackage{graphicx}
\usepackage{textcomp}

\newcommand{\VVSM}{\ensuremath{V}}
\newcommand{\SVSM}{\ensuremath{S}}

\newcommand{\VSMcaln}{\ensuremath{\alpha^\star}}
\newcommand{\JVLF}{\ensuremath{J_\mathrm{VLF}}}
\newcommand{\JMLF}{\ensuremath{J_\mathrm{MLF}}}
\newcommand{\Jc}{\ensuremath{J_\mathrm{c}}}
\newcommand{\Jcmag}{\ensuremath{J_\mathrm{c,m}}}
\newcommand{\Jctrans}{\ensuremath{J_\mathrm{c,t}}}
\newcommand{\Ec}{\ensuremath{E_\mathrm{c}}}
\newcommand{\Emag}{\ensuremath{E_\mathrm{m}}}
\newcommand{\Emaga}{\ensuremath{\bar{E}_\mathrm{m}}}
\newcommand{\Etrans}{\ensuremath{E_\mathrm{t}}}
\newcommand{\Ectrans}{\ensuremath{E_\mathrm{c,t}}}

\newcommand{\fref}[1]{Fig.~\ref{#1}}
\newcommand{\Fref}[1]{Figure~\ref{#1}}
\newcommand{\eref}[1]{Eqn.~\ref{#1}}
\newcommand{\Eref}[1]{Equation~\ref{#1}}
\newcommand{\SUST}[1]{{\it Supercond.~Sci.~Technol.}}
\begin{document}

\title{Magnetic measurement of the critical current anisotropy
in coated conductors}

\author{F. Hengstberger, M. Eisterer, H. W. Weber}
\email{hengstb@ati.ac.at}
\affiliation{Atominstitut,
Vienna University of Technology,
Stadionallee 2,
1020 Vienna}

\begin{abstract}
  We determine the critical current anisotropy at maximum Lorentz force
  from hysteresis loops in a vibrating sample magnetometer.
  To eliminate the signal of spurious variable Lorentz force currents
  it is sufficient to cut the sample to a specific length,
  which is calculated from the position dependent sensitivity of the instrument.
  The procedure increases the resolution of the measurement
  and the results compare well to transport data on the same sample.
  As the electric field in magnetisation measurements
  is lower than in transport experiments
  the anisotropy at high currents
  (low temperatures and fields)
  can be measured
  without the need of making current contacts
  or any special sample preparation.
\end{abstract}


\maketitle

\section{Introduction}

Current transport is the principal application of superconducting films
and it is therefore straightforward to carry out transport measurements
of the critical current density (\Jc),
but certain limitations make an alternative characterisation method desirable.
First,
the resistive heat produced by the current contacts
inhibits the measurement of the very high critical currents
occuring at low temperatures and fields.
Second,
the electric field is rather high
due to the limited voltage resolution on short test samples
and the conductor is therefore even more prone to thermal instabilities.
Magnetisation measurements solve the above problems:
the experiment is contact-less and the electric field,
which is defined by the sweep rate of the applied magnetic field,
is about one order of magnitude lower than in transport measurements
(see below).

\begin{figure}
  \centering
  \includegraphics[width=.2\textwidth]{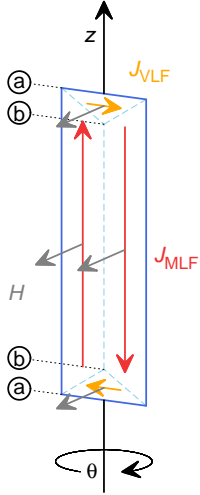}
  \caption{\label{fig:geometry}
    Geometry of the measurement.
    The magnetic field $H$ is perpendicular to the $z$-axis,
    which is the direction of the sample motion in the VSM
    and the axis of rotation
    ($\theta=0^\circ$ if $H$ is perpendicular to the film).
    The extended Bean model accounts for
    two different $a$,$b$-plane current densities
    flowing under maximum Lorentz force (\JMLF)
    along the tape
    or under variable Lorentz force (\JVLF)
    when closing the loop at the end.
    The labels mark the $z$-range,
    where the VLF-currents flow
    (see \fref{fig:coils}).}
\end{figure}

Due to the geometry of magnetisation measurements
(see \fref{fig:geometry})
it is,
however,
impossible to obtain the \Jc{} anisotropy
under the same well defined conditions as in transport experiments.
If a conductor is rotated around its long axis
in a magnetic field applied perpendicular to the axis of rotation,
the induced currents flow in the $a,b$-plane,
which is a consequence of the thin film geometry,
but they flow under different forces:
the currents directed along the tape flow
always at right angles to the magnetic field---%
a configuration identical to transport measurements at maximum Lorentz force (MLF);
the currents closing the loop at the end of the tape
flow under variable Lorentz force (VLF).
As a consequence,
the tape carries two different critical current densities,
which cannot be extracted from the measurement of a single quantity,
i.e.,
the magnetic moment of the tape.

These spurious VLF-currents must therefore be eliminated
to substitute transport by magnetisation measurements.
One possibility is to pattern the conductor into many strips,
which increases the length-to-width aspect ratio
and decreases the contribution of the VLF-currents to the magnetic moment.\cite{Tho10}
Apart from the additional experimental effort
the striation reduces the total magnetic moment of the sample
$m \propto w^2/n$
($w$ is the original width, $n$ the number of cuts)
at the expense of the resolution of the measurement.

In the following we will show that it is sufficient
to cut the conductor to a certain length
when measuring in a transverse vibrating sample magnetometer (VSM).
The method is simple and increases the resolution of the measurement.

\section{Method}

\begin{figure}
  \centering
  \includegraphics[width=.5\textwidth]{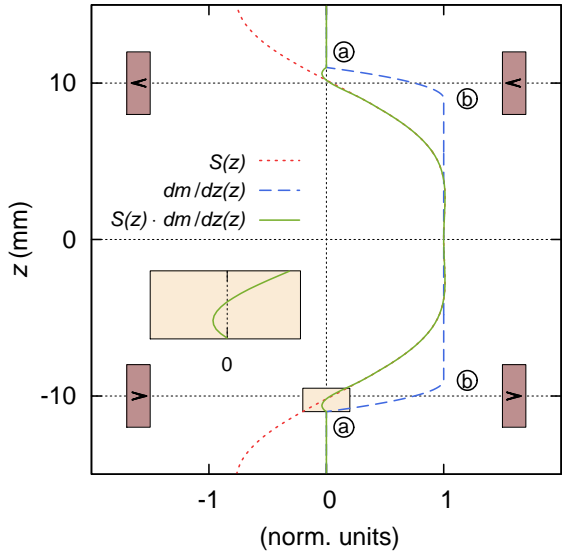}
  \caption{\label{fig:coils}
    A Mallinson coil-set consists of two pairs of coils (brown rectangles)
    with opposite winding direction (indicated by the arrows).
    The VSM sensitivity $\SVSM(z)$ is positive between the coils
    and changes sign approximately at the $z$-position
    of the pick-up coils.
    The VLF-currents start to flow
    where $dm/dz(z)$
    decreases to zero.
    (Confer \fref{fig:geometry} for the label positions.)
    If this area of the sample is at a position,
    where $\SVSM(z)$ is small and changes sign,
    the contribution of the VLF-currents to the total signal,
    which is proportional to the area under $\SVSM(z)\cdot{}dm/dz(z)$,
    is drastically reduced.
    The inset is a magnification of the area,
    where the VLF-currents close the loop.    
  }
\end{figure}

Our approach is based on the fact that the sensitivity of a VSM
equipped with a Mallinson coil set~\cite{Mal66}
(the standard pick-up coil geometry)
depends on the position of the magnetic moment.
For our approach it is sufficient
to take only the $z$-dependence into account.
The convolution of the line density of magnetic moments
$dm/dz(z)=\int\!\!\int dxdy\,M(x,y,z)$
along the $z$-dimension of the sample
with the VSM sensitivity function $\SVSM(z)$
is the VSM output signal 

\begin{equation}
  \VVSM(z)
  =
  \int dz'\, \SVSM(z')\, dm/dz(z-z')\;.
  \label{eqn:vsm}
\end{equation}

(Here, $z$ refers to the distance between the centre of the sample
and the centre of the coil set.)
If a small sample
($dm/dz(z)=m\,\delta(z)$ for a magnetic dipole)
is scanned along the $z$-axis,
$\VVSM(z)\propto\SVSM(z)$:
the symmetric function is positive
approximately up to the position of the pick-up coils
and becomes negative outside
(see \fref{fig:coils}).
Measuring a small sample at the centre position
determines the calibration constant $\alpha=1/\SVSM(0)$
in $m=\alpha\,\VVSM(0)$.

If the $z$-extension of the sample can't be neglected,
the convolution~\Eref{eqn:vsm} comes into effect,
the above calibration is invalid
and we have to distinguish between
the real magnetic moment $m_\mathrm{r}=\int dz\,dm/dz(z)$ of the sample
and the magnetic moment \emph{sensed} by the instrument
$m_\mathrm{s}=\alpha\VVSM(0)$.
On the other hand,
we can take advantage of the $z$-dependence of the VSM sensitivity:
if the spurious VLF-currents at the end of the film
are close to the pick-up coils,
where $\SVSM(z)$ is \emph{small and changes sign},
their net contribution to $m_\mathrm{s}$ can be eliminated
(see \fref{fig:coils}).
At the same time the signal of the MLF-currents
and the resolution of the measurement increase,
because the currents span the region with positive $\SVSM(z)$ between the coils.

The optimal sample length (defined below)
and the new calibration constant \VSMcaln,
are available from~\eref{eqn:vsm}.
The sensitivity function $\SVSM(z)$
can be calculated~\cite{Lac95} or measured (see above);
integrating the magnetisation $M(x,y,z)$ of a film with two constant critical current densities %
(\JMLF{} and \JVLF{} in the extended Bean model sketched in \fref{fig:geometry})
across the width and the thickness results in $dm/dz(z)$.

We define the optimum sample length as the length,
where the \emph{sensitivity} of the VSM to VLF-currents becomes minimal.
For quantification we consider two limiting cases:
the \emph{sensed} magnetic moment when the field is applied perpendicular
and parallel to the film plane.
In the first case both current densities are equal,
but for all other directions of the magnetic field
\JVLF{} will exceed \JMLF{}.
We assume the worst case and let the ratio $R=\JVLF/\JMLF$ diverge,
if the field is in-plane and the VLF-currents are force free.
Evaluating the upper limit of the relative measurement error

\begin{equation}
  \label{eqn:error}
  \epsilon(l)
  =
  |m_\mathrm{s}(l,R=1)-m_\mathrm{s}(l,R\rightarrow\infty)|/m_\mathrm{s}(l,R=1)
\end{equation}

as a function of the sample length $l$
quantifies the sensitivity of the VSM to changes in \JVLF{}.

\begin{figure}
  \centering
  \includegraphics[width=.5\textwidth]{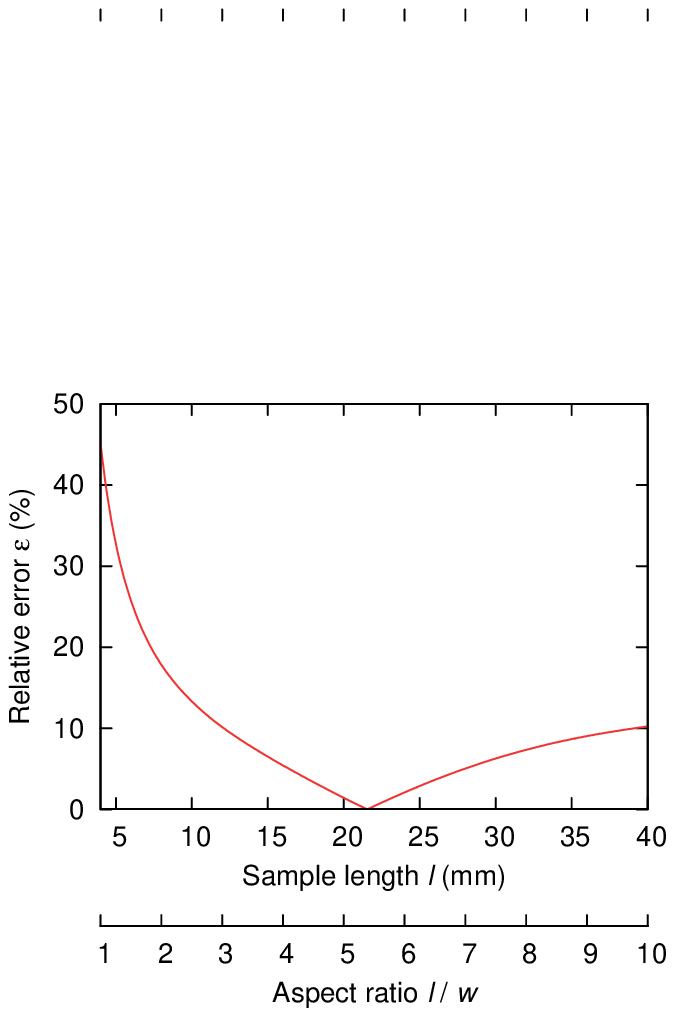}
  \caption{\label{fig:length}
  Calculated relative measurement error as a function of the sample length.
  Cutting the sample to the optimum length reduces the error to below \mbox{1\,\%}
  and makes the VSM insensitive to changes of the VLF-currents.}
\end{figure}

For a \mbox{4\,mm} wide coated conductor we find an optimum length of \mbox{22\,mm}
after calculating $\SVSM(z)$ from theory
and fitting the positions of the pick-up-coils
to a measurement of $\VVSM(z)\propto\SVSM(z)$ in our
Oxford Instruments MagLab VSM using a small calibration sample.
The maximum relative error is below \mbox{1\,\%} at the optimum length
(see \fref{fig:length})
and the influence of the VLF-currents
can be disregarded in the evaluation of \JMLF,
because the VSM is now \emph{insensitive} to changes of \JVLF.

We determine the anisotropy $\Jc(\theta)$ at maximum Lorentz force
from the width of the hysteresis
using the signals of two coil-sets
parallel and orthogonal to the direction
of the applied magnetic field

\begin{equation}
  \label{eqn:moment}
  m_\mathrm{s}=\sqrt{\Delta m_{\mathrm{s},\mathrm{p}}^2+\Delta m_{\mathrm{s},\mathrm{o}}^2}\;.
\end{equation}

The evaluation is only valid
if the currents induced by the last change of the applied field
have fully penetrated the sample.
Since the penetration field of a thin film
$B^\star\propto \Jc\,d/\!\cos(\theta)$
scales with the field normal to the sample~\cite{Mik04},
this criterion cannot be fullfilled for the entire angular range.
Depending on the critical current of the conductor
and the maximum field of the VSM magnet
a certain region close to the $a$,$b$-planes 
(cf. \fref{fig:cpeak})
remains inaccessible.
This is the only limitation of the method.

The simultaneous measurement of magnitude and direction
of the magnetic moment in~\Eref{eqn:moment}
is an important advantage.
If the instrument is equipped with only one parallel coil-set,
the direction of the magnetic moment must be known
to determine $m_\mathrm{s}=m_{\mathrm{s},\mathrm{p}}/\!\cos(\theta)$.
In this case a small angular misaligment will
introduce an asymmetry in the anisotropy curve.
Moreover,
even the smallest alignment error is strongly amplified
close to $\theta=90^\circ$,
where both $\cos(\theta)$ and $m_{\mathrm{s},\mathrm{p}}$ go to zero.

\section{Results}

\begin{figure}
  \centering
  \includegraphics[width=.5\textwidth]{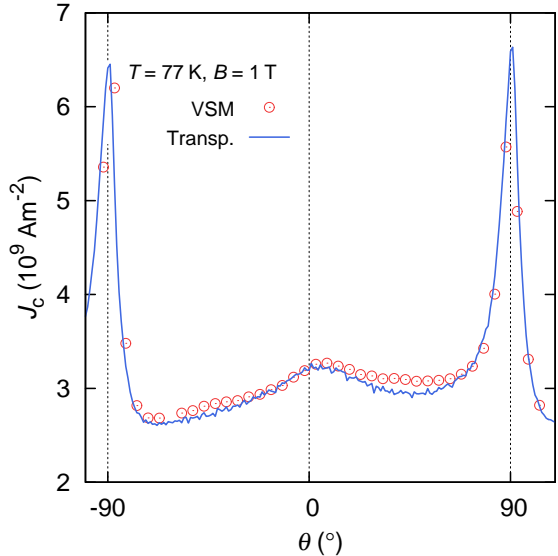}
  \caption{\label{fig:comp}
  Comparison of magnetisation and transport measurements.
  The anisotropy measured in the VSM agrees well with transport data,
  showing a narrow $a,b$-peak and a broad maximum around the $c$-axis,
  features peculiar to coated conductors.
  Note the significant asymmetry in both anisotropy curves.}
\end{figure}

The corresponding experiments were made by performing hysteresis loop measurements
at constant angles in the VSM
(55\,Hz vibration frequency, 0.05--0.15\,mm amplitude)
sweeping the magnetic field at a rate of
$\mu_0dH/dt = 0.5\,\mathrm{T}/\mathrm{min}$.
A 1\,\textmu$\mathrm{V}/\mathrm{cm}$ electric field criterion
defined the critical current in the four-probe transport measurements.
Both experiments were carried out on the same sample,
a YBCO coated conductor grown by MOCVD on a non-magnetic Hastelloy substrate
(the single YBCO layer is approximately 1\,\textmu{}m thick).
The sample is 4\,mm wide and was cut to the optimal length of 22\,mm
(see Fig.~\ref{fig:length})
after the transport measurements,
which require a slightly longer sample length (about 3\,cm)
for low-resistance current contacts.

\Fref{fig:comp} demonstrates that our method
agrees very well with transport measurements on the same sample.
The two main features of a coated conductor's anisotropy,
a sharp $a$,$b$-peak and a broad $c$-axis maximum,
are almost identical.
Note,
that also the asymmetry of the transport anisotropy curve is reproduced.
The magnetisation measurement displayed in \fref{fig:comp}
is calibrated against the transport measurement at $\theta=0^\circ$.
Although the calibration constant differs by only \mbox{20\,\%}
from the calculated \VSMcaln{},
the deviation is significant and will be discussed below. 

\begin{figure}
  \centering
  \includegraphics[width=.45\textwidth]{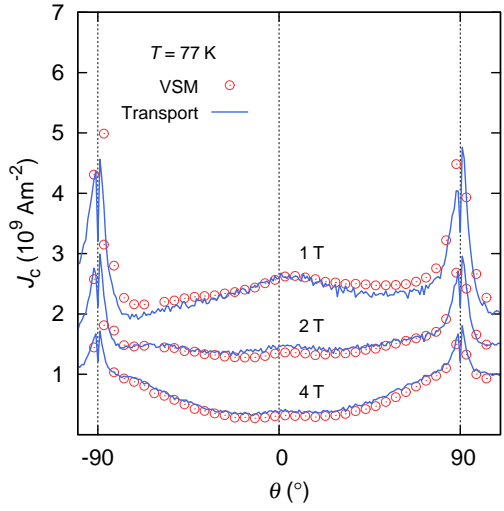}
  \caption{\label{fig:field}
  Extrapolation of the transport anisotropy curve
  to the electric field level of the VSM magnetisation measurement.
  When measurements at different magnetic fields are compared,
  the electric field must be considered.
  Extrapolating the transport \mbox{$I$-$V$ curves}
  to the low electric fields of a magnetisation measurement
  leads to satisfactory agreement over a large field range.
  Note,
  that the asymmetry observed at \mbox{1\,T} disappears at higher fields.}
\end{figure}

When comparing measurements at different magnetic fields
it is important to take the electric field dependence of the critical current $\Jc(E)$
into account,
because the $n$-value of the $E(J)=\Ec\cdot(J/\Jc)^n$ power-law
decreases strongly with increasing magnetic field~\cite{Tho08}.
Without accounting for this well-known effect
the different electric field levels in both instruments
would lead to a magnetic field dependent calibration constant.

A rough estimate of the electric field in the magnetisation measurement
is provided by integrating Faraday's law
in a cylindrical coordinate system
aligned with the applied magnetic field.
The field sweep induces an azimuthal electric field 
$\Emag=r/2\cdot\mu_0 dH/dt$,
which ranges from zero in the centre of the sample to approximately
\mbox{0.1 -- 0.5\,\textmu$\mathrm{V}/\mathrm{cm}$} at the edges of the
$22 \times 4\,\mathrm{mm}^2$ conductor,
showing that the average electric field in a magnetisation measurement
is significantly below the transport criterion
$\Ectrans=$\mbox{1\,\textmu$\mathrm{V}/\mathrm{cm}$}.
(The indices $t$ and $m$ denote transport and magnetisation in the following.)

We account for the different electric fields
by approximating the transport \mbox{$I$-$V$ curves}
with a power-law $\Etrans(J)=\Ectrans\cdot(J/\Jctrans)^n$ 
and extrapolating the transport data
to the average electric fields \Emaga{} of the magnetisation measurements
$\Jcmag=\Jctrans\cdot[\Etrans/\Emaga\cos(\theta)]^{1/n}$.
(The additional factor of $\cos(\theta)$
stems from the angle dependent change of flux trough the sample.)
After fitting \VSMcaln{} and \Emaga{}
the measurements compare well
over a large field and angular range
(see \fref{fig:field}),
except,
of course,
for parallel fields,
where $\cos(\theta)=0$ and the electric field breaks down.
Satisfactory agreement between transport and magnetisation measurements
of the critical current anisotropy in superconducting thin films has,
to our knowledge,
not been published so far.

The average electric field $\Emaga=0.1$\,\textmu$\mathrm{V}/\mathrm{cm}=\Etrans/10$,
is well within the range of our previous estimation.
Taking the electric field into account
reduces also the deviation of \VSMcaln{} from theory
to below \mbox{10\,\%}.
The remaining error can be attributed to uncertainties
in the superconducting sample dimensions
and the fact
that especially currents close to the edge of the conductor
contribute differently to critical current
and magnetic moment.

We wish to emphasise that the low electric field
in magnetisation measurements is certainly not a disadvantage.
The \mbox{1\,\textmu$\mathrm{V}/\mathrm{cm}$}-criterion in transport measurements
is rather a concession to the voltage noise on short test samples
than a technologically relevant criterion---%
in most applications coated conductors will operate at electric fields
much below this criterion.
From this perspective,
magnetisation experiments are superior,
because they are able to explore large critical currents
with high resolution down to very low electric fields.

\begin{figure}
  \centering
  \includegraphics[width=.5\textwidth]{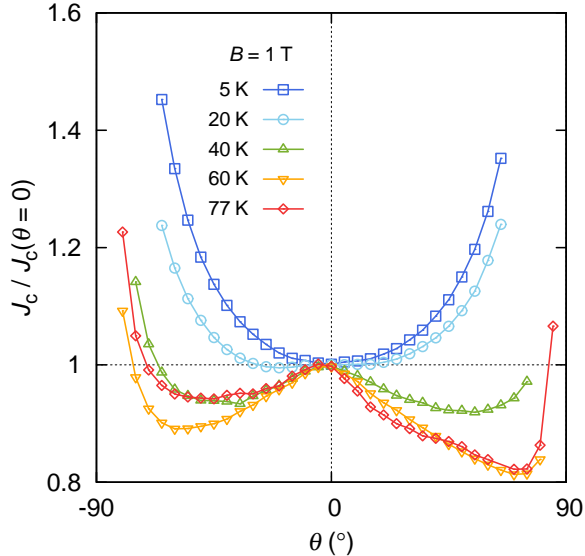}
  \caption{\label{fig:cpeak}
  Disappearance of the $c$-axis peak at low temperatures.
  The $c$-axis peak
  is observed down to relatively low temperatures
  (about \mbox{20\,K} at \mbox{1\,T})
  and vanishes at \mbox{5\,K}.
  The curvature of the anisotropy curve at $\theta=0^\circ$
  thereby changes sign.
  The increase of the penetration field
  limits the angular range at low temperatures.}
\end{figure}

As an example we analyse the temperature dependence
of the $c$-axis peak in coated conductors down to
temperatures as low as 5\,K.
This has hitherto been impossible
due to the power dissipation in transport measurements
and was only recently~\cite{Xu10} achieved at 4.2\,K
directly in liquid helium.
Measurements at 1\,T (see \fref{fig:cpeak})
show that the $c$-axis peak vanishes at 5\,K.
This behaviour is identical to the transport measurements mentioned above,
which were carried out on different samples:
the anisotropy curve showed a $c$-axis peak at 77\,K,
but there was no indication of this feature at 4.2\,K.
The entire temperature dependence,
has,
to our knowledge,
not been reported so far.
We are able to monitor the evolution of the $c$-axis peak with our method
and find that correlated pinning effects shape the anisotropy
down to \mbox{20\,K} at \mbox{1\,T},
but not at lower temperatures.

\section{Concluding remarks}
The method described in this work is derived for the thin film geometry
and may thus be applied not only to coated conductors
but to any superconducting thin film of appropriate dimensions.
In the case of a magnetic substrate the background signal has to be subtracted,
for example,
by removing the superconducting layer
or by measuring a piece of substrate with identical dimensions.
This procedure is,
however,
only necessary below the saturation field of the magnetic substrate,
because according to (\ref{eqn:moment})
a reversible background cancels in the evaluation of \Jc{},
which depends only on the irreversible magnetic moment.

In general we expect small differences between the magnetic and transport measurements
at low applied fields,
i.e.,
when the self-field
(roughly 200\,mT at 5\,K for our sample)
of the currents is similar to the applied magnetic field.
In this case the direct or indirect (via the magnetic substrate)
interaction between the self-field and the field dependent critical currents
will differ between transport~\cite{San10} and magnetisation,
because the field profile of the circulating induction currents 
is different from that of the transport currents.
The self-field regime is,
however,
not important for most applications,
which require a detailed knowledge of the critical current anisotropy.

\section{Summary}

We have shown that
the critical current anisotropy of a coated conductor at maximum Lorentz force
can be measured in a transverse vibrating sample magnetometer.
Simply cutting the sample to a defined length
eliminates the contribution of spurious variable Lorentz force currents
to the magnetic moment sensed by the instrument.
The results obtained in this way compare well to transport experiments on the same sample
if the effect of the different electric field,
which is below the resolution of transport measurements,
is accounted for by extrapolating the transport \mbox{$I$-$V$ curves}.

Although the large penetration fields inhibit measurements
of the critical current close to the film plane,
the advantages of magnetisation measurements,
i.e.,
the lack of current contacts
and the low electric field,
are obvious.
The experiment is thus particularly suited
to explore the critical current anisotropy
at low magnetic fields and temperatures,
which remains inaccessible to transport measurements
due to thermal problems.


\end{document}